# Online Decomposition of Surface Electromyogram into Individual Motor Unit Activities Using Progressive FastICA Peel-off

Haowen Zhao, Xu Zhang, Maoqi Chen and Ping Zhou

*Abstract*—Surface electromyogram (SEMG) decomposition provides a promising tool for decoding and understanding neural drive information non-invasively. In contrast to previous SEMG decomposition methods mainly developed in offline conditions, there are few studies on online SEMG decomposition. A novel method for online decomposition of SEMG data is presented using the progressive FastICA peel-off (PFP) algorithm. The online method consists of an offline prework stage and an online decomposition stage. More specifically, a series of separation vectors are first initialized by the originally offline version of the PFP algorithm from SEMG data recorded in advance. Then they are applied to online SEMG data to extract motor unit spike trains precisely. The performance of the proposed online SEMG decomposition method was evaluated by both simulation and experimental approaches. It achieved an online decomposition accuracy of 98.53% when processing simulated SEMG data. For decomposing experimental SEMG data, the proposed online method was able to extract an average of 12.00 ± 3.46 MUs per trial, with a matching rate of 90.38% compared with results from the expert-guided offline decomposition. Our study provides a valuable way of online decomposition of SEMG data with advanced applications in movement control and health.

*Index Terms*—Surface electromyography, motor unit, online decomposition, progressive FastICA peel-off

## I. Introduction

Electromyogram (EMG) is an electrophysiological signal generated by muscular activation, reflecting motor control commands of the neuromuscular system [1]. It can be used to analyze movement behaviors, intentions and health [2]-[4]. Surface EMG (SEMG) refers to the EMG signals recorded by electrodes placed on the skin surface. Due to its noninvasive manner, SEMG has been widely applied in human-machine interfaces [5]-[7], sports medicine [8]-[9] and rehabilitation [10]-[12]. Ideally, an EMG signal is composed of multiple action potentials generated by activated motor units (MUs), transmitted and superimposed temporally and spatially at a recording electrode [13]. Specifically, each MU consists of the cell body and dendrites of an alpha motor neuron, the multiple branches of its axon, and the muscle fibers that are innervated [14]. The MU is regarded as the basic component of the peripheral neuromuscular system to describe the neural control of muscular contraction and movement formation [15]. Compared with the global features such as SEMG amplitude, the MU activities can reflect the information of neural drives to the muscle at a microscopic level. Therefore, it is valuable to examine the MU activities and properties. EMG decomposition enables resolving the composite EMG signal into its constituent MU spike trains (MUSTs) and MU action potential (MUAP) waveforms. The availability of these individual MU activities can provide a promising way of decoding motor neural commands of a neurophysiological nature [16]-[22].

Many efforts have been made toward EMG decomposition, mainly relying on blind source separation (BSS) algorithms which are aimed to solve the difficult math problem of separating sources from observed signals without prior knowledge of the source signals [23]. Besides, it brings huge challenges to the SEMG decomposition due to its special characteristics such as low signal-to-noise ratio, high similarity and severe superposition of the MUAP waveforms, caused by the low-pass filtering effect of the subcutaneous skin and fat tissues. With the recent development of electronic and sensing technologies, the use of high-density SEMG (HD-SEMG) by 2-dimensional flexible electrode arrays provides abundant spatial information simultaneously recorded from dozens or even hundreds of SEMG channels, facilitating implementing the BSS algorithms in general, and the SEMG decomposition in particular [24]. Convolution kernel compensation (CKC) [25] and progressive FastICA peel-off (PFP) [26] are both representative HD-SEMG decomposition methods, inspired by the advanced BSS techniques [23], [27]. The CKC estimates and updates cross-correlation vectors between the observed SEMG signals and MUSTs in an iterative way [23]. The PFP applies a classic FastICA algorithm [27] to the SEMG signals to calculate the separation vectors and introduces a "peel-off" procedure to progressively remove the separated MUAP waveforms from the original SEMG signals. Such a procedure mitigates the effect of the already identified MUs on the

This work was supported by the National Natural Science Foundation of China under Grant No. 61771444.

H. Zhao and X. Zhang are with the School of Information Science and Technology at University of Science and Technology of China, Hefei, Anhui, 230026, China (email: xuzhang90@ustc.edu.cn).

M. Chen and P. Zhou are with Faculty of Biomedical and Rehabilitation Engineering, University of Health and Rehabilitation Sciences, Qingdao, Shandong, 266024, China (email: dr.ping.zhou@outlook.com).

FastICA convergence and effectively increase the number of obtained MUs. The performance of both CKC and PFP has been extensively validated [28]-[32]. Variations of both methods have been developed to extract a relatively large number of MUs at high muscle contraction levels, with successful applications mainly in offline conditions [33]-[37].

Considering the application prospects of SEMG in many fields, there are substantial demands for robust online SEMG decomposition. Glaser et al. [38] conducted a pilot study on the real-time SEMG decomposition based on the CKC algorithm and demonstrated its feasibility. Afterwards, more relevant studies were reported [39]-[44]. The development of these online decomposition algorithms mainly relies on a basic assumption that SEMG signals are quasi-stationary, and the MU behaviors do not change in pattern over a short period of time. This assumption has served as a primary basis of conventional offline SEMG decomposition [25], [26]. On this basis, these online decomposition algorithms were always designed to use results from an offline decomposition as prior knowledge, thus saving computational resources and allowing the feasibility of online signal processing. Specifically, most previous studies conducted online SEMG decomposition using modified versions of the CKC method, whereas the online version of the PFP method has not been investigated yet. Considering the advantages of the PFP method in extracting a great many MUs with high precision, it is necessary and promising to develop its online version.

Accordingly, this paper presents an online SEMG decomposition method based on the PFP algorithm, evolving the key techniques of the PFP algorithm to meet the requirements for its real-time usability. To avoid the time-consuming complexity from the offline decomposition methods, the proposed method utilized a two-stage approach consisting of an offline prework stage and an online decomposition stage. Furthermore, an adaptive threshold selection algorithm was developed to make it more suitable for precisely determining each MUST while processing in real time. The performance of the proposed online decomposition method was validated on both simulated and experimental SEMG datasets.

## II. RELATED WORK

### A. SEMG Observation

Each MU has a unique and stable MUAP waveform distribution pattern in different channels of a 2-dimensional array, which can be used to distinguish and identify the MU. The SEMG signal can be observed by a convolutional mixing model expressed as [45]:

$$x_i(t) = \sum_{j=1}^{N} \sum_{\tau=0}^{L-1} a_{ij}(\tau) s_j(t-\tau) + n_i(t) \quad (1)$$

where $i = 1,2,3 \ldots \ldots M$ and $t = 1,2, \ldots \ldots T$, $x_i(t)$ is the $i$th SEMG channel and $n_i(t)$ represents the additive noise in the $i$th channel. $a_{ij}(\tau)$ denotes the waveform vector of length L, which represents the waveform of the $j$th MU in the $i$th channel. $s_j(t) = \sum_k \delta(t - T_j(k))$ is the MUST expressed as a 0-1 impulse sequence indicating every spike firing timing at $T_j(k)$ for the $k$th firing of the $j$th MU, whereas $\delta$ is Dirac Delta function. For each $k$, $T_j(k+1) - T_j(k) > L$ can be assumed.

Define the expansion vector of EMG signals and MUSTs as $\overline{x}(t) = [x_1(t), x_1(t-1), \ldots, x_M(t), \ldots, x_M(t-K+1)]$ and $\overline{s}(t) = [s_1(t), s_1(t-1), \ldots, s_M(t), \ldots, s_M(t-K+1)]$.

Thus, the equation can be rewritten in matrix form:

$$\overline{x}(t) = \overline{A}\overline{s}(t) + \overline{n}(t) \quad (2)$$

where $\overline{n}(t)$ represents noise. $\overline{A}$ is a matrix containing all waveform vectors $a_{ij}$. For the mixing model analyzed above, the task of EMG decomposition is to find a suitable separation matrix $\overline{W}$ that consists of many separation vectors to extract the MU firing events. As a result, the source signals of all MUs can be estimated by $\hat{s}(t) = \overline{W}\overline{x}(t)$.

### B. Automatic PFP (APFP)

The PFP algorithm has been automated, but it is suitable just for offline data processing. More details of the algorithm and the corresponding parameters can be found in [33] and the APFP method was used in this study with the same settings as reported in [33]. Below is a brief introduction to the APFP method.

If a whitened observed signal $x$ has been obtained and we need to find an independent component $y = w^T x$ from it using the ICA algorithm [23], [27], the following maximum negative entropy problem needs to be optimized:

$$\begin{aligned} \max \quad & J_G(w) = [E\{G(w^T x)\} - E\{G(v)\}]^2 \\ s.t. \quad & h(w) = E\{y^2\} - 1 = \|w\|^2 - 1 = 0 \end{aligned} \quad (3)$$

where $G$ is a non-polynomial function, and $v$ is a random variable with standard normal distribution.

The problem above can be solved using the procedure of the fix-point algorithm [46] to obtain a series of MU source signals and their corresponding separation vectors. The spike trains can be precisely extracted from these source signals using the initial threshold determined by the Otsu algorithm [47]. However, the spikes from one source signal often do not just belong to one MU due to heavy MUAP superimposition or high MU synchronization levels. Thus, a valley-seeking clustering approach [48] is used to distinguish the spikes from the same source signal based on their morphological features. On this basis, the spikes belonging to each cluster are most likely from the same MU [33]. After the valley-seeking clustering approach, the constrained FastICA algorithm [49] is performed using the extracted and clustered spike trains as constraints to converge. Therefore, the MU source signals can be effectively updated and meanwhile the possible firing errors are corrected. To assess the reliability of the constrained FastICA outputs and their corresponding MUSTs representing true MU activities, some metrics are employed from the perspective of the significance of correlation constrain [49], including the consistency of spike amplitudes and inter-spike intervals [50], and the physiologically reasonable firing rate [51]. In the APFP method, the correlation coefficient between the output of constrained FastICA and the testing spike trains (denoted as $\xi$), the coefficient of variation of spike amplitudes and inter-spike intervals (denoted as $CoV_{amp}$ and $CoV_{isi}$), and the firing rate

(denoted as *FR*) are employed. Moreover, a two-step criterion describing a reasonable range of the above four metrics is employed to judge the MU reliability comprehensively [33].

A "peel-off" procedure is performed later to subtract the obtained MUAP waveforms from the original signals. The MUAP waveforms of the identified MUs were estimated by a straightforward approach following a least squares problem [26], [52] instead of the conventional high-resolution alignment algorithm [53]. More MUs can emerge when processing the residual signals again with the FastICA algorithm. The framework of the offline APFP method is summarized as follows:

(1) Initialize the residual signal to the original EMG signal, and make the MUST set γ empty.
(2) Apply the FastICA algorithm to the expanded residual signal and obtain a series of source signals.
(3) Extract non-repetitive spike trains by Otsu algorithm and use valley-seeking clustering to distinguish these spikes to separate spike trains from different MUs.
(4) Use MUSTs obtained in step (2) as a reference signal, and apply the constrained FastICA algorithm on the expanded original EMG signal to detect the reliability of the MUSTs and to correct possible erroneous or missing discharges.
(5) Judge whether the MUs obtained are reliable through metrics calculation. Put reliable results in set γ.
(6) Estimate the waveforms of the reliable MUs, subtract the estimated MUAP waveforms from the original signal and update the residual signal.
(7) If no new reliable MU is found in the above steps, or the APFP method reaches the preset termination condition, the algorithm ends. Otherwise, go back to step (2).

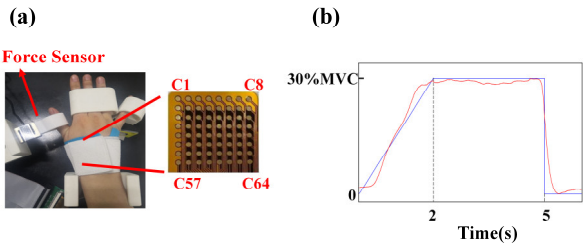

Fig. 1. The experimental setup and protocol. (a) Apparatuses for simultaneously recording thumb abduction force by a load cell and HD-SEMG data by a piece of 2-dimensional electrode array arranged in an 8×8 formation. (b) The illustration of the force generation pattern with both the designed force curve (blue line) and an actual recorded force curve (red line) in one trial of task performance.

## III. METHODOLOGY

### A. Experimental SEMG Data Collection and Preprocessing

#### 1) Subjects and Experiments

Eight subjects (26.13±4.29 years) without any known history of muscular or neural disorder participated in this study. The study was approved by the Ethics Review Board of the University of Science and Technology of China (Hefei, China). All subjects signed consent prior to any procedure of the experiments.

In this work, the HD-SEMG data were recorded from abductor pollicis brevis (APB) muscle due to its wide explorations and applications in SEMG studies [19]-[21]. Here, a home-made, multi-channel signal acquisition system with a force sensor and a set of 3D-printed apparatuses was used to collect data, as shown in Fig. 1a. The subject's hand was placed on the fixed 3D-printed apparatus to prevent muscular movement interferences from the wrist and other fingers, and the muscle force was recorded by a load cell (LDST-V-HY, Luckly Inc., Beijing, China) connected to a ring around the thumb. Multiple electrodes were arranged in the form of 8 rows × 8 columns to form a 2-dimensional electrode array. Each electrode probe had a diameter of 2 mm, and the inter-electrode distance between consecutive electrodes was 4 mm. Each electrode was designed in a monopolar manner relative to a round common reference electrode placed on the back of the tested hand.

During the experiments, subjects were asked to sit and place the tested hand in a relaxed and comfortable way. Before data collection, the maximum voluntary contraction (MVC) of the thumb abduction muscle was tested and recorded. Then, in each trial of the task performance, subjects were instructed to perform isometric muscle contractions with the muscle force gradually increasing from 0 to a targeted force level (quantified by MVC percentage) in 2s and then maintained at the targeted level for around 3s, as shown in Fig. 1b. According to this force generation pattern, the designed force curve was shown on the screen to facilitate the subject's task performance in each trial. The targeted force level in this experiment was set to 30% MVC and the trial was repeated at least nine times to acquire a sufficient amount of data. The force and SEMG data were digitized via a 16-bit A/D converter (ADS1198, Texas Instruments, TX) at a sample rate of 2 kHz, and the data were stored into the hard disk of a computer and imported into the MATLAB software (version R2020a, MathWorks, Natick, MA, USA) for further analyses.

#### 2) Data Preprocessing

All channels of the recorded HD-SEMG signals were inspected, and a few channels (3.75 ± 1.28 channels across all subjects in this study) with low quality were discarded (due to their excessive noise contamination resulting from motion artifacts, occasional electrode drop, or environmental interferences from surrounding electronic devices). The channel deletion remained consistent within the EMG signals of the same subject. The HD-SEMG signals within the remaining channels were filtered through a 10-order Butterworth band-pass filter to reduce possible low-frequency or high-frequency interference. The bandwidth of the filter was 20-500Hz. Finally, the power line interference was removed through a 50Hz second-order notch filter. The deleted channels were not considered in the subsequent process of SEMG decomposition, but they were filled in by interpolation from neighboring channels and considered during the estimation of MUAP waveforms. In order to facilitate the data analysis, all of the SEMG data were divided into a series of non-overlapping data segments corresponding to the force generation task repetitions over time. Therefore, the length of every SEMG data segment was around 5 seconds.

TABLE I
PARAMETERS FOR SEMG SIMULATION

| | Distribution | Mean | SD | Range |
|---|---|---|---|---|
| Fiber number | Uniform | 70000 | | ±0.5 mean |
| MU fiber endplate center position | Uniform | 0 | | ±8 mm |
| Fiber endplate position variation | Uniform | 0 | | ±2 mm |
| Half fiber length | Gaussian | 40mm | 4mm | ±2 SD |
| Mean fiber diameter for a MU | Gaussian | 55μm | 10μm | ±2 SD |
| Fiber diameter variation within a MU | Gaussian | 0 | 1μm | ±2 SD |
| ISI variation | Gaussian | 0 | 0.2*instant mean ISI | ±2 SD |

*B. SEMG Data Simulation*

A data simulation approach was conducted to generate HD-SEMG data with known MU activities, which were used as the ground-truth for validating the performance of the developed online SEMG decomposition method. In the current study, this approach was based on simulation models well described by previous studies, including the motoneuron pool model [54], the model describing the MUAP waveforms of different MUs, and a tripole model [55] considering the generation and extinction of the action potentials at the fiber end-plate and tendon.

Here a cylindrical muscle with a radius of 8 mm was simulated and the fat and skin layers of the muscle were set to 2.5 mm thickness. 120 MUs were set and distributed in parallel in the muscle fibers. Most of the MUs had low recruitment thresholds and a few had high thresholds. When the excitation exceeded the threshold, every MU discharged at 8 Hz and its firing rate increased as the excitation increased. All the relevant parameters are listed in Table I.

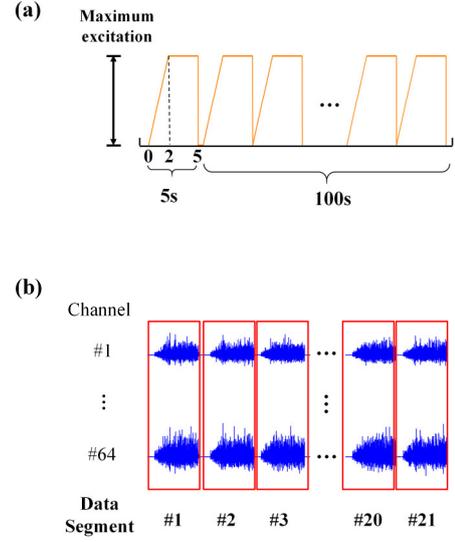

Fig. 2. (a). The contraction condition of simulated signals. (b). Multi-channel simulated SEMG signals.

The simulated SEMG signals were also set to be recorded by a 64-channel surface electrode array arranged in an 8×8 grid form. The inter-electrode distance was set at 4 mm for both horizontal and vertical directions. The electrode array was placed parallel to the muscle fiber direction and its center electrodes were set to approximately over the innervation zones.

To be consistent with the force generation pattern of the actual experiments, the excitation was set to increase from 0 to a specific excitation level in the first 2 seconds, and maintained for another 3 seconds with several repetitions. The maximum excitation level was set to be 3%, corresponding to 33 active MUs. In addition, zero-mean Gaussian noises were added to the simulated EMG signals, generating three levels of SNR (signal-to-noise ratio) at 10 dB, 20 dB and 30 dB, respectively. Thus, we considered four noise levels, three SNR levels and the level without any additional noise. For each noise level, 21 repetitions were simulated to ensure data diversity, as shown in Fig. 2. Therefore, 84 data segments (4 noise levels × 21 repetitions) were simulated in total.

*C. Online Decomposition*

The overall whole block diagram summarizing the proposed online decomposition method is described in Fig. 3.

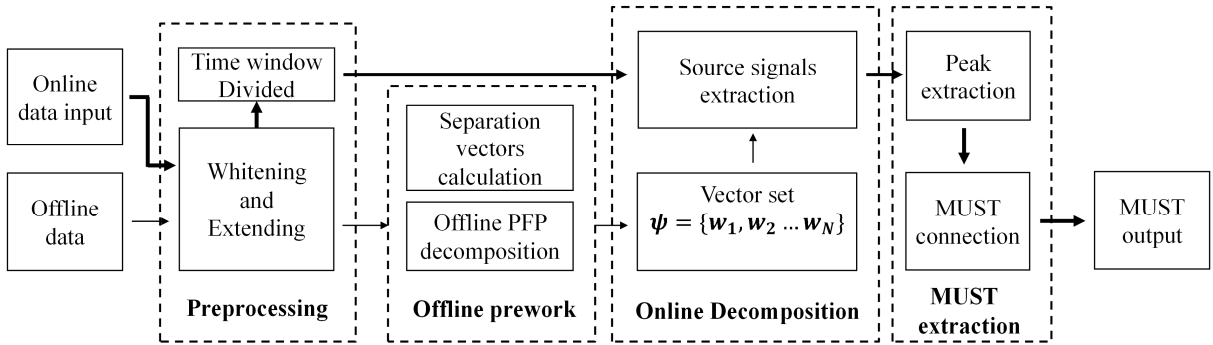

Fig. 3. Block diagram of the proposed method for online SEMG decomposition

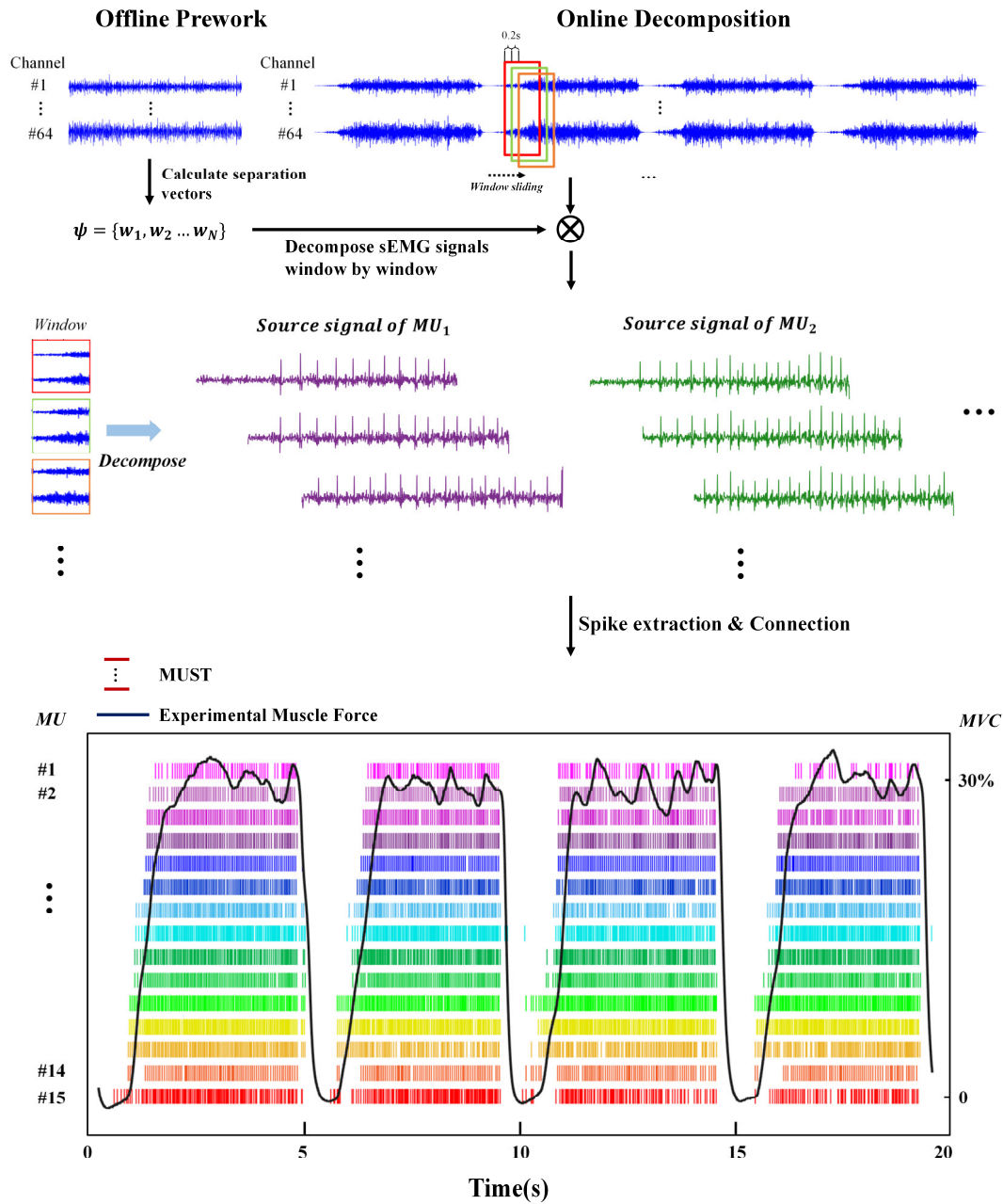

Fig. 4. Illustration of the online SEMG decomposition process using the proposed method.

With full consideration of the real-time usability of the proposed online method, a two-stage approach was designed to avoid considerable computational complexity caused by the repeated operation of the FastICA algorithm and the iterations of the constrained FastICA algorithm. More specifically, the reliable separation vectors were initialized in the offline prework stage and saved to accelerate the subsequent online data processing. In the online decomposition stage, the data stream of the input EMG signals was divided into a series of temporally overlapping windows with window length and increment set at 1 and 0.2 s, respectively. Both settings helped to facilitate online processing.

During the offline prework stage, several 5-s segments of EMG signals were separately decomposed offline using the APFP method and all of the resultant separation vectors were put into the set $\psi$. The quality of these vectors was evaluated by both criteria employed in the offline APFP method [33]: if the coefficient of variation of spike amplitudes $CoV_{amp}$ was higher than 0.3, and the coefficient of variation of inter-spike intervals $CoV_{isi}$ was higher than 0.4, the corresponding separation vector was considered to be low-quality and it was removed from the set $\psi$. Furthermore, any duplicated separation vector corresponding to the same MU was removed as well.

In the online decomposition stage, every 0.2 s of data input was combined with 0.8 s of historical data to form a 1-s window for decomposition. The decomposed results from consecutive windows were connected, while their overlapping portion was used to align the obtained MUSTs. This ensured continuity of

**Algorithm 1** The proposed online decomposition method

1: Decompose the SEMG signals offline. Extract the MUSTs and calculate the corresponding separation vectors.
2: Remove the duplicated separation vectors and vectors that are not well-decomposed.
3: Save all the separation vectors $w_1, w_2, w_3 \ldots w_N$ for the online decomposition stage.
4: **while** Acquiring SEMG signals **do**
5:     Load and extend the EMG signals ($\bar{y}$).
6:     **for** $j = 1; j < N + 1; j ++$ **do**
7:         Calculate the source signal, $s_j = w_j^T \bar{y}$.
8:         Estimate the initial threshold through the Otsu algorithm and extract the spike train.
9:         Successively increase the threshold and extract a series of spike trains $ST_{j1}, ST_{j2}, ST_{j3}\ldots$
10:         Find the spike train with the lowest $CoV_{amp}$ and $CoV_{isi}$ as the $j$th MUST $ST_j$.
11:     **end for**
12:     Connect the MUSTs over the sliding windows.
13: **end while**

the decomposition results along with the original SEMG data stream. The SEMG data in each window were first whitened and extended. Then, the multiplication procedure was directly applied to the extended EMG signals with separation vectors in set $\psi$ to estimate different MU source signals, from which individual MUSTs were consequently identified.

For extracting MUSTs from the MU source signals, the original offline APFP method employs repeated iterations of the constrained FastICA algorithm, involving complex computations as described above. This process was unsuitable for online processing and therefore it was removed to avoid heavy computational burden. To maintain high-precision MUST extraction, the simple amplitude-thresholding process by the Otsu algorithm had to be updated. A new algorithm was designed for our online PFP method. First, this algorithm needs to determine an initial threshold that is applied to each source signal, using the Otsu algorithm in the same way as conducted in the offline APFP method. Then, a group of spikes beyond this threshold is detected and the corresponding amplitudes can be ranked from small to large. Next, a series of successively increasing thresholds that are a little higher than these amplitudes are adopted to estimate a series of different spike trains. Each resultant spike train can be further evaluated by both $CoV_{amp}$ and $CoV_{isi}$ metrics, and the spike train with the minimal summation of both metrics is finally considered the most appropriate MUST. This algorithm for adaptive threshold selection was termed the successive multi-threshold Otsu algorithm.

A k-means clustering algorithm was usually used in some offline decomposition methods [36]-[37] for extracting MUSTs from the source signals. It was also implemented in this study as an alternative threshold selection algorithm, in comparison to the successive multi-threshold Otsu algorithm used in our method. By applying the k-means clustering algorithm, all sample amplitudes of the source signal time series can be classified into 2-4 groups (2 in this work), so that the group with the largest amplitudes of samples is selected as the extracted MUST.

After the spike trains of all MUs were appropriately detected, they were connected over windows to form the resultant MUST for each MU, and its MUAP waveforms that spanned over all channels were correspondingly estimated. Fig. 4 illustrates an example of the online decomposition results. The pseudocode of the proposed online decomposition method is shown in Algorithm 1.

*D. Performance Evaluation*

For processing the experimental SEMG data, the proposed online decomposition method was conducted in a user-specific manner. Four segments were used in the offline prework stage and the remaining 4 segments were processed in the online decomposition stage. For processing the simulated SEMG data, the first segment was used in the offline prework stage and the remaining 20 segments were processed in the online decomposition stage. All SEMG segments tested in the online decomposition stage were sequentially arranged in the form of a data stream to be processed continuously using our proposed method. For comparison purposes, all of the SEMG segments to be processed online was also decomposed by the offline APFP method as well.

To evaluate the performance of online decomposition and assess the decomposition results more comprehensively, we

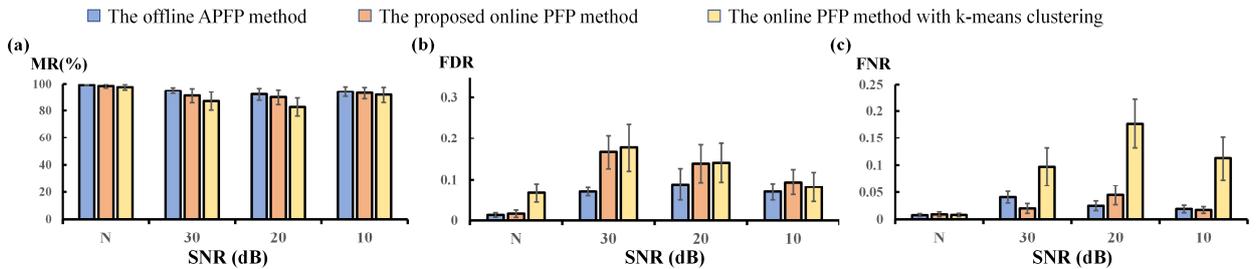

Fig. 5. The results for decomposing simulated SEMG data in terms of MR(a), FDR(b) and FNR(c) averaged over all data segments using the offline APFP method, the proposed online PFP method and the online PFP method with k-means clustering at four noise levels, respectively. The error bar represents standard deviations. N in the horizontal axis denotes the condition without any additional noise.

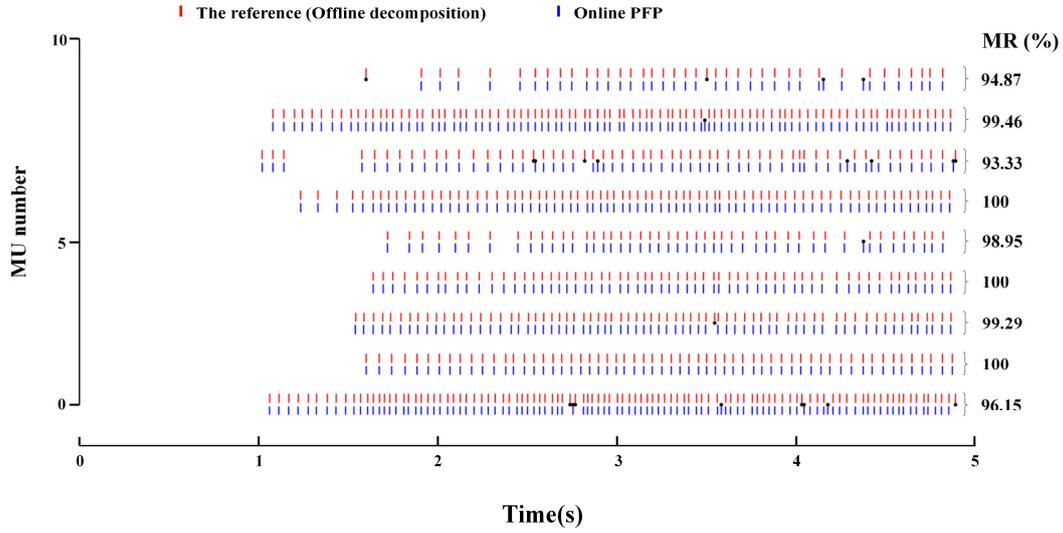

Fig. 6. A representative example of validating the decomposition results from the online PFP method in terms of all decomposed MUSTs (in blue) with respect to the reference (in red) derived from summarized offline decomposition results, using a data segment from one subject. The position of the black dot indicates the missing or fault discharges and MR values are computed and shown on the right side of these spike trains.

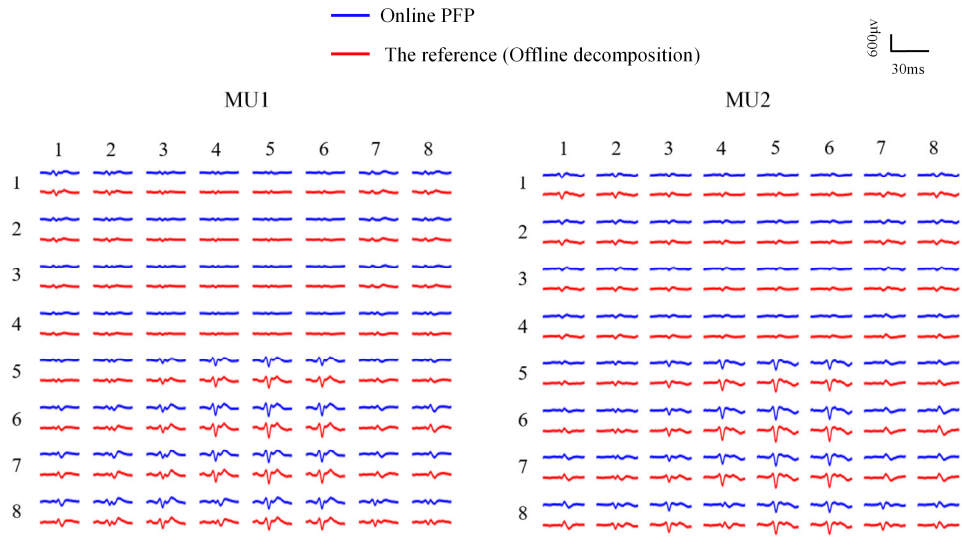

Fig. 7. Two MUAPs of matched MUs with time-varying waveform shapes. Here we illustrate 64 electrode channels arranged in an 8×8 grid form. Blue and red lines indicate the MUAP shapes from online PFP and the reference of offline decomposition, respectively.

used a series of metrics: matching rate (MR) can be calculated as [33]:

$$MR = \frac{2 \cdot N_{common}}{N_{online} + N_{reference}} \quad (4)$$

where $N_{online}$ denotes the number of firing events of the online decomposition results, and $N_{reference}$ denotes the number of the reference spike trains. In the simulated data, the reference spike train indicates the ground-truth firing events. However, the actual MUSTs are not known a priori in the experimental data. Therefore, the decomposition results of the experimental data processed by the offline APFP method were used to define $N_{reference}$. $N_{common}$ indicates the number of common discharges appearing in both the online decomposition result and the reference. The MR measures the matching degree and it is able to quantify the precision of an online decomposition method.

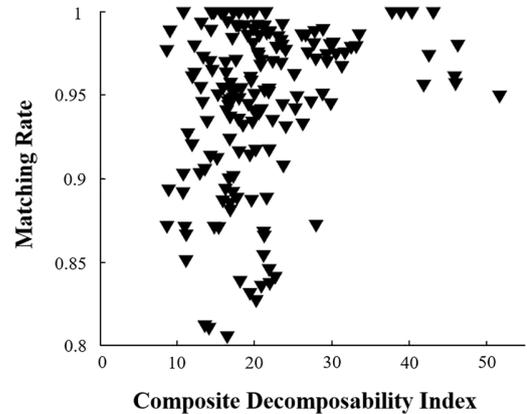

Fig. 8. The relationship between the matching rate and the composite decomposability index.

Besides MR, both false negative rate (FNR) and false discovery rate (FDR) were used to reveal the cause of the error discharges. They are defined as

$$FNR = \frac{N_{reference} - N_{common}}{N_{reference}} \quad (5)$$

$$FDR = \frac{N_{online} - N_{common}}{N_{online}}$$

They count the proportion of the number of unmatched discharges to the total number of their respective discharges. Specifically, the FNR measures the rate of "missing" discharges with respect to the reference, and the FDR quantifies the rate of "faulty" discharges appearing in the online decomposition results. Generally speaking, the MR of a reliable MUST is close to 1 but the FNR and FDR are close to 0.

For a more comprehensive view of the decomposition results, we also calculated the mean discharge rate (MDR) and the coefficient of variation (CoV) of the online identified MUSTs with respect to the reference spike trains. It should be noted that the CoV refers to the coefficient of variation of the inter-spike intervals $CoV_{isi}$ to better understand the MU firing behaviors.

In addition, we calculated the decomposability index (DI) for each common MU of experimental EMG data to precisely quantify the proposed method's performance [56]:

$$DI = \frac{\min\{\|m_{ki}\|, \|m_{ki} - m_{k*i}\|\}}{V_i^{RMS}} \quad (6)$$

where $m_{ki}$ is the MUAP of the $k$th MU in the $i$th channel and $m_{k*i}$ is the MUAP most similar to $m_{ki}$ among the other MUAPs in the $i$th channel. $V_i^{RMS}$ is the root mean square amplitude (RMS) of the $i$th channel and the operator $\|\cdot\|$ denotes the Euclidean norm. The DI measures the separation between $m_{ki}$ and the template of MUAP nearest to it (or the baseline), normalized by the standard deviation of the noise component (interference plus baseline noise) projected along their vector difference. The overall decomposability of the $k$th MU was measured by the composite DI (CDI), defined as the norm of the individual DIs [56].

For developing a real-time decomposition method, it is necessary to evaluate the processing time delay which is expected to be as short as possible. The time delay for processing one single time window was recorded, and all these time delay values were averaged across all windows and all subjects to indicate the computational complexity. All of the algorithms were implemented on a desktop computer with an Intel Core i5-10400 processor (2.90 GHz) and 16 GB of memory.

## IV. RESULTS

### A. Results of Simulated Data

As an offline decomposition method for validation, 21 MUs were identified using offline APFP and the number was 22 using online PFP when no additional noise was added. Further, the number of MUs correctly decomposed using online PFP decreased to 11, 7, and 6 when noise was added at 30 dB, 20 dB and 10 dB SNR, respectively.

The results for decomposing simulated SEMG data are reported in Fig. 5. As compared with the offline APFP method, the proposed online PFP method achieved comparable performance in terms of a high MR over 90%, and a low FNR below 0.05. The proposed online PFP method had a fluctuated and relatively higher FDR than the offline APFP method under three SNR levels. Specifically, a decreasing trend of the MR was found from 99.29% to 94.13% for the offline APFP method and from 98.53% to 92.79% for the online PFP method, respectively, when the noise was successively added to generate four noise levels. The ANOVAs revealed no significant difference in either MR, FDR or FNR, between the offline APFP method and the proposed online PFP method ($p > 0.05$).

When both threshold selection algorithms were compared, it was evidently found that the successive multi-threshold Otsu algorithm in the proposed online PFP method significantly outperformed the K-means clustering algorithm in terms of higher MR ($p = 0.025$) and lower FNR ($p = 0.022$). Both algorithms did not exhibit a significant difference in the FDR metrics ($p = 0.273$).

Table II reports both MDR and CoV values calculated for all common MUs between the decomposition results achieved by the proposed online method and the ground truth. The ANOVA revealed no difference in MDR ($p = 0.217$) or CoV ($p = 0.105$) at no presence of noise. However, the MDR and CoV of online decomposition results became significantly different from those of the ground-truth ($p < 0.05$) when the noises were added.

TABLE II
COMPARISON OF MDR AND COV OF THE SIMULATED EMG SIGNALS

|  | SNR 10dB Online PFP/ Ground-truth | SNR 10dB Online PFP/ Ground-truth | SNR 30dB Online PFP/ Ground-truth | No adding noise Online PFP/ Ground-truth |
|---|---|---|---|---|
| MDR | 9.86±1.99 8.77±0.18 | 9.55±1.54 8.75±0.23 | 10.47±1.81 8.74±0.22 | 8.77±0.51 8.70±0.18 |
| CoV | 0.245±0.053 0.199±0.003 | 0.257±0.032 0.202±0.005 | 0.231±0.044 0.201±0.005 | 0.211±0.024 0.199±0.007 |

### B. Results of Experimental Data

When implementing online decomposition of experimental data, the offline decomposition method was applied to establish the reference for validation, and 10.31±1.79 MUs were obtained, averaged across all subjects.

Fig. 6 is an example of an online decomposition result using the proposed method, showing the decomposed MUSTs with respect to the reference. It can be observed that almost all the MU discharges derived from the online PFP method are well matched with those in the reference, with sporadic missing or erroneous ones. Fig. 7 illustrates the MUAP waveforms of two matched MUs derived from both the online PFP method and the reference, which demonstrate a very consistent waveform shape in each channel and almost the same distribution pattern across the electrode array. Fig. 8 plots the relationship between the matching rate and composite decomposability index (CDI), which displays the overall trend of the matching rates varying

TABLE III
SUMMARY OF DECOMPOSITION RESULTS FOR EXPERIMENTAL EMG SIGNALS.

| Subject | Number of motor units | | | MDR (Hz) | | CoV (%) | | MR (%) | FDR | FNR |
|---|---|---|---|---|---|---|---|---|---|---|
| | The reference | Online PFP | | The reference | Online PFP | The reference | Online PFP | | | |
| | | All | Matched | | | | | | | |
| 1 | 12.75±1.50 | 19 | 9.00±1.41 | 19.76±5.08 | 19.71±4.43 | 22.92±7.71 | 24.31±8.53 | 92.06±5.91 | 0.084±0.082 | 0.051±0.057 |
| 2 | 9.50±1.29 | 8 | 4.50±0.58 | 22.00±4.76 | 20.63±4.00 | 27.44±6.59 | 22.46±5.95 | 89.92±7.21 | 0.093±0.075 | 0.106±0.086 |
| 3 | 9.00±0.82 | 14 | 6.00±0.81 | 15.79±3.22 | 14.65±3.30 | 23.44±3.78 | 25.44±4.59 | 93.20±6.02 | 0.065±0.068 | 0.067±0.074 |
| 4 | 11.00±1.41 | 13 | 8.75±1.71 | 20.15±3.93 | 21.28±3.84 | 24.08±6.98 | 24.67±6.25 | 91.17±3.35 | 0.076±0.033 | 0.056±0.029 |
| 5 | 8.50±0.57 | 9 | 5.50±0.58 | 20.29±3.99 | 20.62±3.00 | 26.09±4.19 | 28.48±4.70 | 85.18±4.04 | 0.116±0.051 | 0.175±0.068 |
| 6 | 9.50±1.29 | 10 | 6.25±1.71 | 20.35±4.25 | 19.67±4.30 | 23.87±3.05 | 24.18±3.73 | 91.51±6.45 | 0.084±0.071 | 0.082±0.076 |
| 7 | 11.75±1.71 | 11 | 7.00±0.82 | 23.03±3.60 | 24.66±3.94 | 24.46±3.54 | 24.82±2.78 | 87.26±5.47 | 0.131±0.073 | 0.108±0.028 |
| 8 | 10.50±1.29 | 12 | 6.50±1.73 | 18.57±2.72 | 18.73±1.86 | 18.74±2.96 | 19.41±1.66 | 92.70±4.26 | 0.080±0.058 | 0.064±0.040 |
| Average | 10.31±1.79 | 12.00±3.46 | 6.69±1.84 | 19.99±2.18 | 19.99±2.79 | 23.88±2.55 | 24.22±2.56 | 90.38±2.80 | 0.091±0.022 | 0.089±0.041 |

with the CDIs. It contains the common MUs of all of the collected SEMG segments.

Table III reports both the number of MUs decomposed by the online PFP method and the number of common MUs matched those in the reference (offline decomposition) for 8 subjects, respectively. An average of 12.00±3.46 MUs were successfully identified by the online PFP method, with an average of 6.69±1.84 MUs correctly matched. Besides, three metrics are also computed from those common MUs and reported in Table III. Averaged over all data segments to be decomposed and all subjects, the MR was (90.38±2.80) %, the FDR was 0.091±0.022, and the FNR was 0.089±0.041. The estimated MDR ($p = 0.872$) and CoV ($p = 0.503$) of online decomposition results were not significantly different from the offline decomposition reference.

*C. Time Delay*

The time delay for decomposing a 1-s window of SEMG data using the proposed method in the online decomposition stage was 0.084±0.028 s, averaged over all data segments and all subjects; it was less than a 0.2-s time increment. For comparison purposes, the offline APFP method costs 60.07 ± 9.82 s to decompose SEMG data in a single time window, much longer than that of the proposed online decomposition method.

V. DISCUSSION

As a promising SEMG decomposition method, the PFP algorithm has been reported recently and, therefore, it is necessary and promising to develop an online version. This study sought to propose an online SEMG decomposition method based on the PFP algorithm. The results of both simulated and experimental SEMG data analyses demonstrated the feasibility of the proposed online PFP method in decomposing a large number of MUs with high precision in the context of isometric muscle contractions. Our study offers a valuable tool for online SEMG decomposition with great applications in biomechanics and rehabilitation.

In the results of processing simulated data, the proposed online PFP method decomposed a similar number of MUs as the offline APFP method, illustrating comparable performance. Due to the use of initial separation vectors provided by the APFP method in the offline prework stage, the proposed online PFP method is expected to inherit a good capability of decomposing a great number of MUs from its original offline version. In terms of MR, the proposed online PFP method got a slightly lower value compared with the offline APFP method. This can be explained by the fact that the source signals were calculated by directly multiplying previously initialized separation vectors with the SEMG signals for the purpose of real-time processing. In addition, the MUSTs were estimated without the examination of iterative constrained FastICA, thus increasing the negative influence of noise. The result demonstrates that online decomposition was speeded up at the cost of a little bit of decrease in precision. This is the main and common difficulty in generalizing an offline decomposition method to its online version [38]-[43]. However, it has been found that the MDR and CoV of online decomposition were significantly different from those of the ground-truth when the noise was added. This can be partly explained by the limitations of the online decomposition method such as MU synchronization [26] and firing events drift [33] that previous studies have faced.

When some noises were successively added to EMG signals to be decomposed, both the number of correctly identified MUs and the precision of determining their firing timings were reported to decrease substantially. This could partly explain that the decrease of SNR resulted in more serious noise interference to some small MUs and thus caused a negative influence on the calculation of separation vectors as well as the performance of the online decomposition method. On the other hand, it became much harder to precisely extract MUSTs from source signals in the online decomposition stage at a low SNR level, reflecting the decline of the MR. As a consequence, it can be inferred that the quality of SEMG signals significantly influenced the performance of the decomposition method, as reported in [33], [40].

It is worth mentioning that the proposed online PFP method introduced a progressive multi-thresholding process for extracting MUSTs. The successive multi-threshold Otsu algorithm outperformed the conventional k-means clustering algorithm especially in the condition of noise interference, proving the potential to extract more precise discharges at low SNR levels. The successive multi-threshold algorithm based on the Otsu algorithm was inspired from the common Otsu algorithm [48] used in the offline APFP method [33]. It was able to successively increase multiple thresholds to overcome the effect of noise interferences and find the most appropriate one to extract MUSTs that followed the physiological properties of MUs. The successive multi-threshold Otsu algorithm takes consideration into the interval and waveform information to ensure the result to be much more reliable, depending on $CoV_{amp}$ and $CoV_{isi}$. By contrast, the k-means clustering algorithm only focuses on the amplitude information of EMG source signals. As a result, it makes it much more difficult to remove the noise interferences and leads to decomposition performance degradation. The proposed online PFP method replaced the complex iterative calculation of constrained FastICA with the successive multi-threshold Otsu algorithm to extract MUSTs, showing a significant improvement in reducing the calculation complexity while maintaining its high precision.

To evaluate the real-time performance, this study recorded the processing time of online decomposition. The time delay was effectively reduced from 60 seconds for the offline APFP method to less than 0.08 seconds for the online decomposition. The acceleration of data processing is attributed to reasons in two respects. The first is that the repeated iteration of FastICA was put in the offline prework stage, which initialized the separation vectors for online decomposition. On the other hand, some complex calculation procedures were adaptively simplified. For example, the constrained FastICA algorithm in the APFP method was replaced with the successive multi-threshold algorithm, as discussed above.

In the experimental SEMG data, a large number of MUs decomposed by offline PFP can be correctly identified with high precision in the online decomposition process, demonstrating that the separation vectors used in the online decomposition process were comprehensive and precise. In addition, the MDR and CoV of online decomposition showed no significant difference with the offline reference. These findings indicate that the performance of the online decomposition method is very close to that of the original offline method, proving the feasibility and effectiveness of the proposed online PFP method. In addition, it illustrates that the advantages of the offline APFP method were still maintained in the proposed online decomposition method.

There are still some limitations in this work. First, the online decomposition process relied too much on the separation vectors provided by the offline prework, proving the feasibility that the separation vectors obtained from offline decomposition can be used for online decomposition. However, the conditions of muscle contraction change over time and the initialization process needs to update the separation vectors, which has not been validated in this work. In other words, the online process was verifying whether the MUs corresponding to the separate vectors were activated and the newly recruited MUs couldn't be captured. Moreover, the initial MU information and spike drift needs to be corrected over time. Second, the experimental EMG data were collected only from isometric contraction and most muscle contractions in daily life are non-isometric and dynamic. More contraction patterns will be added to the experimental data for analysis. Third, the peel-off procedure needs to be adopted in a real-time way to find more MUs and fully take advantage of the offline PFP method. Further research will be devoted to overcoming the limitations above.

## VI. CONCLUSION

A new online SEMG decomposition method based on the Progressive FastICA Peel-off procedure was proposed in this paper, including offline prework and online decomposition process. The proposed decomposition method took advantage of offline PFP algorithms and demonstrated high precision with the most identified MUs both on simulated and experimental EMG signals. These results offer a new tool for precisely identifying individual MU activities in a real-time way with the potential applications of high-density EMG as a neural interface in the fields of biomechanics, sports and rehabilitation.